\documentstyle[preprint,eqsecnum,aps,epsfig]{revtex}
\tightenlines
\def\graphicwidth{0.6\textwidth}
\def\L{{\cal L}}
\def\half{{1\over 2}}
\def\integ{\int_{-\half}^{\half}}
\def\inth{\int_{-\half}^{\half}}
\def\beq{\begin{eqnarray}}
\def\endeq{\end{eqnarray}}
\def\nn{\nonumber \\}
\def\Romnum#1{\uppercase\expandafter{\romannumeral #1}}

\def\ep{\epsilon}
\def\suma#1{\sum_{#1=0}^{\infty}}

\makeatletter
\newtoks\@stequation

\def\subequations{\refstepcounter{equation}%
  \edef\@savedequation{\the\c@equation}%
  \@stequation=\expandafter{\theequation}
  \edef\@savedtheequation{\the\@stequation}
  \edef\oldtheequation{\theequation}%
  \setcounter{equation}{0}%
  \def\theequation{\oldtheequation\alph{equation}}}

\def\endsubequations{\setcounter{equation}{\@savedequation}%
  \@stequation=\expandafter{\@savedtheequation}%
  \edef\theequation{\the\@stequation}\global\@ignoretrue}
\makeatother

\begin{document}
\draft
\preprint{HUEAP-012,hep-th/9903133}
\title{A New Basis Function Approach to 't~Hooft-Bergknoff-Eller 
Equations}
\author{Osamu Abe\thanks{e-mail address: 
osamu@asa.hokkyodai.ac.jp}
}
\address{
Laboratory of Physics\\ 
Asahikawa Campus, Hokkaido University of Education\\
9 Hokumoncho, Asahikawa 070-8621, Japan}
\date{March 17, 1999}
\maketitle
\centerline{(Revised June 24, 1999)}
\begin{abstract}
We analytically and numerically investigate the 
't~Hooft-Bergknoff-Eller equations,
the lowest order mesonic Light-Front Tamm-Dancoff equations for 
$\rm U(N_C)$ and $\rm SU(N_C)$ gauge theories. We find the 
wavefunction can be well approximated by new basis functions and 
obtain an analytic formula for the mass of the 
lightest bound state. Its value is consistent with the precedent
results.
\end{abstract}
\vspace{7mm}

\section{Introduction}
Light-front (LF) quantization is believed to be an effective method
for studying many-body relativistic field theories 
\cite{brodsky98,burkardt}. The physical vacuum is equivalent to the 
bare vacuum in the LF coordinate, since all constituents must have 
non-negative longitudinal momenta defined by
$k^+=(k^0+k^3)/\sqrt{2}$.  This simple structure of the true vacuum 
enables us to avoid the serious problems which appeared in the 
Tamm-Dancoff (TD) approximation \cite{tammdancoff} in the equal time 
frame. Therefore, the TD approximation \cite{lftd} is 
commonly used in the context of the LF quantization.

The techniques have been developed \cite{ma,mo,harada94} for solving 
LFTD equations in several models such as the massive Schwinger model
\cite{coleman}, which is the extension of the simplest 
(1+1)-dimensional QED${}_2$ \cite{schwinger}.
Bergknoff \cite{bergknoff} first applied LFTD 
approximation to the massive Schwinger model. 
He obtained the so-called Bergknoff equation, which is the light front 
Einstein-Schr\"odinger equation truncated to one fermion-antifermion 
pair. He obtained excellent results for the lowest energy meson under
't~Hooft's ansatz. He also discovered that it is necessary to include 
two fermion-antifermion pairs in order to study the excited states.

Ma and Hiller \cite{ma} studied the lowest Bergknoff equation 
numerically. They developed their numerical method for solving the 
Bergknoff equation using an idea based upon 't~Hooft's ansatz.

Mo and Perry \cite{mo} suggested that even in the first excited state, 
most of the wavefunction consists of four fermion sectors when the 
fermion mass reaches zero. Therefore, four or more fermion sectors 
must be included so as to describe wavefunctions beyond the ground 
state bosons. 

In the bibliography \cite{mo}, Mo and Perry presented an effective way 
to treat the ground state and the excited state in the massive 
Schwinger model. They concluded 
that to study the massive Schwinger model, the Jacobi polynomials are 
suitable as basis functions. Harada and his coworker \cite{harada94} 
studied the massive Schwinger model with $\rm SU(2)$ flavor symmetry, 
including up to four fermion sectors. They used simpler basis 
functions, which are essentially equivalent to the Jacobi polynomials.
Sugihara and collaborators \cite{sugihara} numerically analyzed 
2-dimensional $\rm SU(N_C)$ Quantum ChromoDynamics(QCD) \cite{thooft}, 
including four fermion sectors, by means of the basis functions of 
Harada {\it et al}.

Although excellent papers exist concerning massless and massive 
Schwinger models and 2-dimensional QCD 
\cite{huang,sande,duncan,abe}
, including excited states, 
it is worth analyzing the ``'t~Hooft-Bergknoff-Eller'' equation 
\cite{hornbostel}, the extension of the 't~Hooft-Bergknoff equation, 
in order to include both 
$\rm SU(N_C)$ and $\rm U(N_C)$ gauge theories. 
This is because there is 
a mathematical interest in the basis function method. There is no 
mathematical evidence that the conventional basis function expansion
describes the wavefunction well; instead, the contrary is rather to be 
supposed, as there is evidence that the conventional method 
breaks down if we try to improve the approximation. We would therefore 
like to improve the basis functions so as to avoid such difficulties.

Further, there is a so-called ``2\% discrepancy'' problem, briefly 
summarized as follows:
We expand dimensionless meson mass squared $M^2$ in terms of 
dimensionless quark mass $m$,
\beq 
M^2=1+b_1 m+b_2 m^2 +\cdots,
\endeq
where 
\beq
M^2 ={{\bar M}^2\over \mu(N_C)}\quad {\rm and}\quad
m^2={{\bar m}^2\over \mu(N_C)}.
\endeq
Here, ${\bar M}$ is a mass of bound state, ${\bar m}$ denotes a bare 
mass of quark, and $\mu(N_C)$ stands for 
$\displaystyle \mu(N_C)={({N_C}^2-1+\alpha)g^2\over 2 \pi {N_C}}$.
That is, we measure all masses in the unit of $\mu(N_C)$. 
Banks \cite{banks} {\it et al} obtained first order coefficient $b_1$ 
analytically, using the bosonization method,
\beq
b_1=2e^{\gamma_E}=3.56214\cdots,
\label{boson}
\endeq
where $\gamma_E=0.57721\cdots$ is the Euller's constant. On the other 
hand, Bergknoff \cite{bergknoff} found the value,
\beq
b_1={2\pi\over\sqrt{3}}=3.62759\cdots,
\endeq
which differs from Eq.(\ref{boson}) by 2\%. In \cite{harada98}, the 
authors suggested that the coefficient $b_1={2\pi\over\sqrt{3}}$ was a 
variational invariant and that 
this discrepancy was ascribable to the contributions from the higher 
Fock sectors. Before proceeding on to consider the higher Fock 
sectors, we have to examine whether this discrepancy can be explained 
or not in terms of the lowest light-front Tamm-Dancoff equation using 
all possible basis functions.

The 't~Hooft-Bergknoff-Eller equation for two dimensional gauge theory 
is given in the form 
\beq
M^2\Phi (x) &=& \integ dy H(x,y)\Phi(y)\nn
&\equiv&{4(m^2-1)\over 1-4x^2}\Phi (x)-\wp\integ dy 
   {\Phi (y)\over (y-x)^2} +\alpha\integ dy\Phi (y),
\quad
-\half\le x\le \half,
\label{eller}
\endeq
where $\wp$ denotes the finite part integral, $\alpha=1$ for 
$\rm U(N_C)$, and $\alpha=0$ for $\rm SU(N_C)$.
In Eq.(\ref{eller}), we shifted the variable $x$ total amount of 
$-\half$ compared with the variable in  
\cite{bergknoff,thooft,hornbostel}, in order to show the symmetry 
of the wavefunction transparently.

%
Mo and Perry concluded in \cite{mo} that the Jacobi polynomials, 
$(1-4x^2)^\beta P_n^{\beta,\beta}(2x)$ in our notation, are the most 
suitable basis functions. This conclusion seems quite natural because
the system of the Jacobi polynomials $P_n^{\beta,\beta}(2x)$ is an
orthogonal complete set on the interval 
$-\frac{1}{2} \leq x \leq \frac{1}{2}$ with respect to 
the weight function $(1-4x^2)^\beta$. 
Harada and collaborators \cite{harada94} 
suggested using the simpler basis functions, $(1-4x^2)^{\beta +j}$ 
and $x(1-4x^2)^{\beta +j}$
in our notation, which are equivalent to the abovementioned ones. 
According to Harada {\it et al}, 
one can expect that the wavefunction could be expanded as follows:
\beq
\Phi (x)=\cases{
           \displaystyle\lim_{N\to\infty}
             \sum_{j=0}^{N} a_{j}  (1-4x^2)^{\beta +j},\cr
           \displaystyle\lim_{N\to\infty}
                          \sum_{j=0}^{N} b_{j} x(1-4x^2)^{\beta +j}.\cr}
\label{expansion}
\endeq
Here, we have used the fact, as is shown in Appendix \ref{ap1}, that 
the Eller equation does not mix even and odd functions with each other.

The exponent $\beta$ and the quark mass $m$ are related to each other 
by the  equation \cite{thooft,bergknoff}
\beq
(m^2-1) + \beta\pi\cot\beta\pi =0.
\label{mass}
\endeq
The authors of references  \cite{ma,bergknoff,harada94,sugihara} 
adopted the positive smallest solution $\beta_0(m)$ of Eq.(\ref{mass}) 
as $\beta$ in Eq.(\ref{expansion}). That is, for small $m$, 
\beq
\beta_0(m)={\sqrt{3}m\over \pi}\left(1-{m^2\over 10}\right)
+O(m^5)\equiv {\beta_0}^1m+{\beta_0}^3 m^3+\cdots.
\label{beta0}
\endeq
In the following sections, we try to determine the coefficients $a_n$'s, 
according to the predecessors.

\section{Conventional Basis function Method}
In this section, we restrict ourselves to the case $\alpha=0$ where 
the gauge group is $\rm SU(N_C)$.
Mo and Perry, and Harada and his collaborators, presented effective way 
to determine the coefficients. We will briefly reproduce their 
procedures. By the use of the expansion in Eq.(\ref{expansion}) 
truncated to given finite number $N$ for the wavefunction 
$\Phi $, we multiply both sides of Eq.(\ref{eller}) by 
$(1-4x^2)^{\beta +i}$ and integrate them over x, then we obtain 
\beq
M^2 {\widehat N}\vec{a}={\widehat H}\vec{a},\quad 
\vec{a}={}^t[a_0,a_1,\cdots ,a_{n-1}].
\label{eigen1}
\endeq
Here $\widehat N$ and $\widehat H$ are $n \times n$ matrices and are 
given by
\beq
{\widehat N}_{ij}=\inth dx(1-4x^2)^{\beta +i}(1-4x^2)^{\beta +j}
={\pi^{{1/2}}\Gamma (2\beta+i+j+1)\over 2 \Gamma (2\beta +i+j+{3/2})},
\endeq
and
\beq
{\widehat H}_{ij}&=&
4(m^2-1) \inth dx(1-4x^2)^{2\beta +i+j-1}
-\wp\integ dx dy {(1-4x^2)^{\beta +i}(1-4y^2)^{\beta +j}
        \over (y-x)^2}\nn
&=& 2\pi^{{1/2}}(m^2-1){\Gamma(2\beta +i+j)
        \over \Gamma (2\beta +i+j+{1/2})}\nn
&+&{2^{4\beta +2i+2j-3} (\beta +i)(\beta +j)\over 2\beta +i+j}
B(\beta +i,\beta +i)B(\beta +j,\beta +j).
\endeq
See Appendix of  \cite{harada94}.
So-called ``norm'' matrix $\widehat N$ appeared in the above equation 
because the basis functions we have used are not orthonormalized.
In order to have eigenvalues of the generalized eigenvalue equation 
given in Eq.(\ref{eigen1}), 
we have to solve the eigenvalue problem for norm $\widehat N$ first, 
{\it i.e.},
\beq 
{\widehat N} \vec{v}_i =\lambda_i \vec{v}_i.
\endeq

Next, we introduce a transformation matrix $\widehat W$ by
\beq
{\widehat W}=\left[{\vec{v}_1\over ||\vec{v}_1||\sqrt{\lambda_1}},
    \cdots ,{\vec{v}_n\over ||\vec{v}_n||\sqrt{\lambda_n}}\right].
\endeq
Then, we can transform Eq.(\ref{eigen1}) into a usual eigenvalue 
problem of the form
\beq
M^2 \vec{b}={}^t{\widehat W}{\widehat H}{\widehat W}\vec{b},
\quad \vec{a}={\widehat W}\vec{b}.
\label{eigen2}
\endeq

We can solve Eq.(\ref{eigen2}), numerically. For $N=3$ and $m =0.01$, 
we find, for the ground state boson,
\beq
\beta&=&0.00552328,\quad M^2 =0.0366342,\nn
a_0&=&1,\quad a_1= 0.00203562,\quad
a_2=-0.000579369,\quad 
a_3= 0.000165813.
\label{numeric}
\endeq
The values of the LHS and the RHS of Eq.(\ref{eller}) are shown in 
Fig.~1. The coincidence of the LHS and the RHS is high for small 
values of $x$. For $x\simeq\pm{1/2}$, the behavior of the LHS and the 
RHS are quite different. There are sharp spikes at the end points. 
This behavior is not changed  much even if we improve the order of 
approximation. 

Note here that in order to solve the generalized eigenvalue problem, 
the norm matrix should be positive definite.
We cannot advance the above procedure beyond $N\simeq 12$, because 
some of the eigenvalues of the norm matrix $\hat N$ become almost zero 
or negative.
We will examine mathematically this approximated wavefunction in 
detail in the next section. \\
\fbox{Fig.~1}
\section{An Inspection of the conventional Basis Function Method}
\subsection{Behavior of wavefunction around $x=0$}
We introduce linear map $\L$ by
\beq
{\cal L}: f \mapsto {\cal L}f\quad
{\rm such\quad that}\quad
\left({\cal L}f\right)(x)=\wp\integ dy{f(x+y)\over y^2}.
\endeq
After some tedious but not particularly difficult calculations, we find 
that
\beq
\L: (1-4x^2)^\beta &\mapsto&
-4\pi^{{1/2}}{\Gamma (\beta +1)\over \Gamma(\beta+{1/2})}
 F(1,{1/2}-\beta ;{1/2};4x^2)\nn
\L: x(1-4x^2)^\beta &\mapsto&
-8\pi^{{1/2}}{\Gamma (\beta+1)\over \Gamma(\beta +{1/2})}\,x\,
 F(2,{1/2}-\beta ;{3/2};4x^2).\nn
\label{eq1}
\endeq
where $F(a,b;c;x)$ is the Gauss' hypergeometric 
function \cite{abramowitz}. See Appendix A.

We restrict ourselves to the case where the wavefunction is an even 
function, because we are interested in only the ground state meson. 
Thus, we are led to 
\beq
& &
\lim_{N\to \infty}\sum_{n=0}^{N} a_n M^2(1-4x^2)^{\beta +n}\nn
&=&\lim_{N\to \infty}\sum_{n=0}^{N} a_n \left[4(m^2-1)
   (1-4x^2)^{\beta +n-1}+4\pi^{{1/2}}
{\Gamma (\beta +n+1)\over \Gamma (\beta +n+{1/2})}
F(1,{1/2}-\beta -n;{1/2};4x^2)\right].
\label{starteq}
\endeq

Now, we can examine whether the numerical result obtained so far 
satisfies Eq.(\ref{eller}) or not.
We substitute Eq.(\ref{numeric}) into the above equation, and obtain
\beq
LHS(x)=0.0366936 - 0.00101207\,{x^2} - 0.00165597\,{x^4} 
   - 0.00466671\,{x^6} +  O(x^8)
\endeq
and
\beq
RHS(x)=0.0367133 - 0.0038283\,{x^2} + 0.0503141\,{x^4} 
   - 0.210419\,{x^6} +  O(x^8).
\endeq
Thus, the LHS and the RHS coincide, within numerical errors, with each 
other only up to $O(x^0)$. We calculated coefficients $a_n$'s up to 
$n=10$, but the coincidence between the LHS and the RHS is not much 
improved. See Fig.~1.

\subsection{Behavior of wavefunction near $x=\pm \half$}
Bergknoff suggested that the behavior of wavefunction $\Phi$ near end 
points $x=\pm \half$ is important in order to calculate the mass 
eigenvalue $M^2$. In fact, Eq.(\ref{mass}) is derived by demanding
that the most singular part, that is, the coefficients of 
$(1-4x^2)^{\beta-1}$ on the RHS in Eq.(\ref{eller}) must be cancelled, 
as there is no such term on the LHS. According to Bergknoff's 
suggestion, we will examine the behavior of wavefunction near the 
end-points beyond 
$\displaystyle O\left(\left(1-4x^2\right)^{\beta-1}\right)$. 
We set $4x^2 =1-\ep$.
By the use of the identity for the hypergeometric 
functions, that is, Eq.(15.3.6) in \cite{abramowitz}, which is valid 
for $a+b-c\neq \rm integer$, 
we can expand the RHS of Eq.(\ref{starteq}) 
around $\ep =0$ and have 
\beq
LHS=M^2\suma{n}a_n\ep^{\beta +n}.
\label{left}
\endeq
and
\beq
RHS&=&4(m^2-1)\suma{n}a_n \ep^{\beta+n-1}\nn
&+& 4\pi\suma{n}\sum_{j=0}^{n}a_j (\beta+j)
   \cot\left(\pi\left(\beta+j\right)\right)
  {\frac{(-1/2)_{n-j}}{(n-j)!}}\ep^{\beta+n-1}\nn
&-&2\pi^{1/2}\suma{n}\left[\sum_{j=0}^{\infty}a_j
   \frac{\Gamma\left(\beta+j+1\right)(1/2-\beta-j)_{n}}
        {(\beta+j-1)\Gamma(\beta+j+1/2)(2-\beta-j)_n}
   \right]\ep^n.
\label{right}
\endeq

Substituting the numerical solution, which is given in 
Eq.(\ref{numeric}), into Eqs.(\ref{left}) and (\ref{right}), we have
\beq
LHS(\ep)=0.0366342 \,{\ep^{\beta}} + 
  7.45734\times 10^{-5}\,{\ep^{\beta+1}} - 2.12247\times 10^{-5}\,
  {\ep^{\beta+2}} +O({\ep^{\beta+3}})
\endeq
and
\beq
RHS(\ep)=0.551656  - 
  0.525734 \,{\ep^{\beta}} + 2.09072\,\ep -2.08014\,{\ep^{\beta+1}} 
 + O(\ep^{2}).
\label{endpoint}
\endeq
Only $\ep^{\beta-1}$ terms on the LHS and the RHS coincide with each 
other, because we define $\beta$ so that the coefficients of 
$\ep^{\beta-1}$ cancel each other on the RHS. Note that
the LHS does not contain $\ep^n$ terms with non-negative integer $n$, 
while the RHS does. These tendencies are not changed even if we 
calculate $a_n$'s for $n$=3, 6 and 10.

If we rewrite the first two terms of Eq.(\ref{endpoint}) as
\beq
0.551656(1-\ep^\beta )+0.025922\ep^\beta,
\endeq
we can see the origin of the spikes at the endpoints in Fig.~1. The 
spikes arise from the existence of the constant term in the 
wavefunction. The wavefunction $\Phi(x)\equiv  1$ is the exact 
solution of Eq.(\ref{eller}) for $m^2=0$ and $M^2=0$ in $\rm SU(N_C)$. 
Thus, one may expect that the spikes at the end points are closely 
related to the existence of the massless bound state in $\rm SU(N_C)$. 
This is not the case, however, because we can easily see that the 
constant term, in the wavefunction, is allowed if and only if 
$m^2\equiv 0$. We may conclude that the spikes are nothing but the 
artifact which arose from the fact that we have used the improper basis 
functions in Eq.(\ref{expansion}). In fact, we may remove the spikes 
if we use the suitable wavefunction. Refer to the next section.

The basis function given in Eq.(\ref{expansion}) cannot be a good 
mathematical approximation of the true wavefunction.
The reason is as follows: 
If we truncate the series Eq.(\ref{expansion}) to $N$, we expect that 
Eq.(\ref{eller}) holds up to $O(\ep^{\beta+N-1})$. We have only $N+1$ 
parameters $a_1$, $a_2$, $\cdots$, $a_n$ and $M^2$. On the other hand, 
we have $2N$ equations up to $O(\ep^{\beta+N-1})$. 
No consistent solution can exist in general in this case.

The main difficulty comes from the fact that $\cal L$ maps 
$(1-4x^2)^\beta$
not only to the terms $(1-4x^2)^{\beta +j-1}$ with non-negative 
integer $j$ but also to the terms $(1-4x^2)^{j}$. 
One may expect that the above difficulty is avoidable if the 
additional terms $(1-4x^2)^j$ are introduced in Eq.({\ref{expansion}). 
However, we can easily see that coefficients of such terms must 
cancel. By the use of the identity, that is Eq.(15.3.11) in 
\cite{abramowitz}, we have for $n\ge 2$
\beq
{\cal L}:(1-4x^2)^n\mapsto 
   &-&\frac{1}{2(n-1)}\sum_{k=0}^{n-2}\frac{\Bigl(1/2-n\Bigr)_k}
   {\Bigl(2-n\Bigr)_k} (1-4x^2)^k\nn
   &-&\frac{\pi^{1/2}}{\Gamma(1/2-n)}(-1+4x^2)^{n-1}
      \sum_{k=0}^{\infty}\frac{\Bigl(n\Bigr)_k\Bigl(-1/2\Bigr)_k}
      {k!(k+n-1)!}(1-4x^2)^k\times\nn
   && \left[\log (1-4x^2) -\psi(k+1)+\psi(-1/2+k)\right].
\endeq
Here, $\psi(z)$ denotes the digamma function. An analogous formula for 
$n=1$ holds. See Eq.(15.3.10) in \cite{abramowitz}.
If we introduce a term $(1-4x^2)^n$ in Eq.(\ref{expansion}) with 
positive integer $n$, a new singular term like $\ep^{n-1}\log\ep$ 
appears only on the RHS. Note that the exponent $n$ of $\ep$ is the 
same as that of the introduced term. Thus all the coefficients of 
$(1-4x^2)^n$ should be zero.
\section{New Basis Function}
We must notice that there are infinite solutions of Eq.(\ref{mass}) in
addition to the solution given by Eq.(\ref{beta0}). In fact, we see 
that 
\beq
\beta_n(m)&=& n+{1/2}
  - \frac{1}{\left(n+{1/2}\right)\pi^2}
  - \frac{2}{3\left(n+{1/2}\right)^3\pi^4}
  +O\left(\frac{1}{(n+{1/2})^5\pi^6}\right)
\nn
&+& m^2\left[
   \frac{1}{\left(n+{1/2}\right)\pi^2}
   + \frac{1}{3\left(n+{1/2}\right)^3\pi^4}
   +O\left(\frac{1}{(n+{1/2})^5\pi^6}\right)\right] +O(m^4)\nn
&\equiv& {\beta_n}^0+{\beta_n}^2 m^2 + \cdots,\;n=1,2,3\cdots.
\label{betan}
\endeq
From Eqs.(\ref{beta0}) and (\ref{betan}), we are led to
\beq
&&0\ll \beta_n(m)-\beta_0(m)-n<{1/2},\nn
&&0<\beta_n(m)-\beta_k(m)-(n-k)\ll 1.
\endeq
The above relations imply that Eq.(\ref{expansion}) never incorporates 
terms like $(1-4x^2)^{\beta_n+j}$ with positive integer $n$ and 
non-negative integer $j$.

We posit that the wavefunction is given by an infinite series
\beq
\Phi(x)=\lim_{N\to\infty}\sum_{n=0}^{N}\sum_{j=0}^{N-n} {c_n}^j
   \left(1-4x^2\right)^{\beta_n(m)+j}.
\label{newwave}
\endeq
For counting the number of free parameters and the number of 
nontrivial equations, we consider the truncated wavefunction to given 
finite $N$. The truncated wavefunction includes the term 
$(1-4x^2)^{\beta_N}$ and all the other lower order terms.
We require that Eq.(\ref{eller}) should hold up to 
$O(1-4x^2)^{\beta_N-1}$.
For each given value of $m$, the unknown parameters are $M^2$ and 
${c_n}^j$ except for ${c_0}^0\equiv 1 $.
Thus the number of the parameters is $\frac{(N+1)(N+2)}{2}.$
On the other hand we have $\frac{N(N+3)}{2}$ nontrivial equations. 
See Table \ref{table1}.\\
\fbox{Table 1.}\\

The number of parameters is larger than that of non-trivial equations
by 1. Thus, we can solve the equations for ${c_n}^j$ in terms of $M^2$.
Another equation of use to us is obtained by multiplying 
both sides of Eq.(\ref{eller}) by $\Phi (x)$ and integrating them 
over x,
\beq
M^2\integ dx \left|\Phi(x)\right|^2
&=&\integ\integ dx dy \Phi(x)H(x,y)\Phi(y).
\label{consistency}
\endeq

It should be noted here that Eq.(\ref{newwave}) is, mathematically, 
the most general expansion. This means that there is no room to 
introduce any other additional terms like $d (1-4x^2)^\gamma$ for 
$\gamma\neq\beta_n+j$ with non-negative integers $n$ and $j$. If we 
introduce such terms, then the following equality should hold 
\beq
0=4d\left(m^2-1+\pi\gamma\cot(\pi\gamma)\right)(1-4x^2)^{\gamma-1}.
\endeq
This demands that $d\equiv0$.

In the following subsections, we will examine our new basis function 
in detail both analytically and numerically.
\subsection{An analytic approach}
In this subsection, we will restrict ourselves to the case where $N=1$.
Up to $O(\ep^{\beta_1-1})$, we have four equations:
\begin{subequations}
\beq
\ep^{\beta_0-1}&:&0=4(m^2-1) {c_0}^0+4\pi\beta_0\cot 
     (\pi\beta_0){c_0}^0,\label{auto1}\\
\ep^0&:&
 0={c_0}^0\frac{\Gamma(1+\beta_0)}{(1-\beta_0)\Gamma({1/2}+\beta_0)}
     -{c_0}^1\frac{\Gamma(2+\beta_0)}{\beta_0\Gamma({3/2}+\beta_0)}\nn
&&\;\;+{c_1}^0\frac{\Gamma(1+\beta_1)}{(1-\beta_1)\Gamma({1/2}+\beta_1)
        }\nn
&&\;\;+\frac{\alpha}{4}\left[
         {c_0}^0\frac{\Gamma(1+\beta_0)}{\Gamma({3/2}+\beta_0)}
        +{c_0}^1\frac{\Gamma(2+\beta_0)}{\Gamma(\frac{5}{2}+\beta_0)}
        +{c_1}^0\frac{(1+\beta_1)}{\Gamma({3/2}+\beta_1)}\right],
     \label{eqa1}\\
\ep^{\beta_0}&:& {c_0}^0M^2=4{c_0}^1(m^2-1)+4\pi\cot(\pi\beta_0)
       \left[-\frac{{c_0}^0\beta_0}{2}+(1+\beta_0){c_0}^1\right],
     \label{eqa2}\\
\ep^{\beta_1-1}&:& 0=4(m^2-1) {c_1}^0+4\pi\beta_1\cot 
     (\pi\beta_1){c_1}^0.
\label{auto2}
\endeq
\end{subequations}
Equations (\ref{auto1}) and (\ref{auto2}) are automatically satisfied.
Since ${c_0}^0\equiv 1$, we can solve Eqs.(\ref{eqa1}) and 
(\ref{eqa2}) for ${c_0}^1$ and ${c_1}^0$ in terms of $m$, $M$, 
$\beta_0$, and $\beta_1$. 

Now, in order to solve the above equations for $M$, we assume that all 
physical quantities can be expanded in terms of quark mass $m$. That 
is,
\beq
M^2=b_0 + b_1 m + \cdots.
\label{massseries}
\endeq
Thus, we have, up to $O(m)$,
\beq
&&\integ dx|\Phi(x)|^2\equiv<\Phi|\Phi>\nn
&=&1
+m \left[2 {\beta_0}^1 \left( 22 + 
       8 {\beta_1}^0 \left( 1 - 3 \log (2) \right)  - 12 \log (2) + 
       \alpha \left( 2 - 6 \log (2) - 
          {\beta_1}^0 \left( 5 -6 \log (2) \right)  \right)  \right) 
   \right]\nn
&&\times\left[
     3 \left( -2 + \alpha \left( -1 + {\beta_1}^0 \right)  - 
       4 {\beta_1}^0 \right) \right]^{-1} 
   +{b_1}m\left[\frac{2\left(1 -  {\beta_1}^0\right)}
   {2 + \alpha + 4 {\beta_1}^0 - \alpha {\beta_1}^0}\right] ,
\label{norm}
\endeq
and
\beq
&&\integ \integ dx dy \Phi(x)H(x,y)\Phi(y)\equiv <\Phi|H|\Phi>\nn
&=&\alpha
+m\Biggl[-2 \left( 1 + 2 {\beta_1}^0 \right)  
      \left( 3 + {{{\beta_0}^1}^2} {{\pi }^2} \right)  + 
     \alpha \left( 3 \left( -1 + {\beta_1}^0 \right)  + 
        {\left({{\beta_0}^1}\right)^2} \left( 48 + 6 {\beta_1}^0 
     - {{\pi }^2} \right.\right.\nn
&+& {\beta_1}^0 {{\pi }^2} - \left.\left. 36 \log (2) - 
           36 {\beta_1}^0 \log (2) \right)  \right) \Biggr]
\left[3 
     {\beta_0}^1 \left( -2 + \alpha \left( -1 + {\beta_1}^0 \right)  - 
       4 {\beta_1}^0 \right) \right]^{-1}\nn
&+&{b_1}m\left[\frac{2\left(1 -  {\beta_1}^0\right)}
   {2 + \alpha + 4 {\beta_1}^0 - \alpha {\beta_1}^0}\right] ,
\label{expectation}
\endeq

We may expect that coefficient $b_1$ depends on $\alpha$, as both 
coefficients of $m$ in Eqs.(\ref{norm}) and (\ref{expectation}) 
explicitly depend on $\alpha$. However, this is not the case. Indeed, 
from Eq.(\ref{consistency}), we are led to
\beq
M^2&=&\alpha +\left(
{\frac{1}{{\beta_0}^1}} + {\frac{{\beta_0}^1 {{\pi }^2}}{3}}\right)m
+O(m^2)\nn
&=&\alpha+\frac{2\pi}{\sqrt{3}}m+O(m^2).
\label{aprox}
\endeq
Note that we obtained the first line in the above expansion without 
referring to the explicit value of $\beta_1$. We did not reproduce the 
result of Banks {\it et al},  but that of Bergknoff.
The approximated wavefunction in case $m=0.01$ and $N=1$ is shown in 
Fig.~2. The approximation used here is so rough that the coincidence 
of the LHS and the RHS is poor. Nevertheless, the behavior of the RHS 
near the end points is quite calm compared with the 
results of the conventional basis function method. 
The smoothness of the RHS is quite natural. As mentioned previously,
the wavefunction given in Eq.(\ref{newwave}) is the most general.
If we truncate the wavefunction up to order $N$, it becomes smooth. 
So we may also expect the RHS of the 't~Hooft-Bergknoff-Eller 
equation to be smooth.

\fbox{Fig.~2}

We may expect that 
the coincidence of the LHS and the RHS will be improved if the higher 
order terms are included. 
In cases where $N\ge 2$, we cannot treat things analytically.
We will attempt to solve Eq.(\ref{eller}) numerically by the use of 
new basis function in the next subsection.

\subsection{A Numerical approach}
In general, we have 
\beq 
0&=&\left.M^2 \Phi(x)-\integ dy H(x,y)\Phi(y)\right|_{1-4x^2=\ep}\nn
 &=&-4\suma{n}{c_n}^0\left(m^2-1+\pi\beta_n\cot\pi\beta_n\right)
    \ep^{\beta_n-1}\nn
 &&+\suma{n}\suma{j}\left(M^2 {c_n}^j -4(m^2-1){c_n}^{j+1}
   -4\pi\sum_{k=0}^{j+1}{c_n}^k(\beta_n+k)\cot \pi\beta_n
    {\Bigl(-1/2\Bigr)_{j+1-k}\over (j+1-k)!}\right)\ep^{\beta_n+j}\nn
 &&+{\pi^{1/2}\over 2}\suma{n}\suma{j}{c_n}^j\left(
   {4\Gamma(\beta_n+j+1)\over (\beta_n+j-1)\Gamma(\beta_n+j+1/2)}
   -{\alpha\Gamma(\beta_n+j+1)\over \Gamma(\beta_n+j+3/2)}\right)
   \ep^0\nn
 &&+2\pi^{1/2}\suma{k}\left(\suma{n}\suma{j}{c_n}^j
   {\Gamma(\beta_n+j+1)\over (\beta_n+j-1)\Gamma(\beta_n+j+1/2)}\cdot
   {\Bigl(1/2-\beta_n-j\Bigr)_{k+1}\over \Bigl(2-\beta_n-j\Bigr)_{k+1}}
   \right)\ep^{k+1}.
\label{series}
\endeq
Here, $\ep\equiv 1-4x^2$ as before. Of course, the first line in 
Eq.(\ref{series}) cancels automatically because of the definition of 
$\beta_n$'s. Then, suppose that we truncate series in 
Eq.(\ref{newwave}) to $O(\ep^{\beta_N})$. That is, we set ${c_n}^j=0$ 
for $n+j>N$. For a given $m$, we put $M^2=M_i^2$.
We can then solve Eq.(\ref{series}) for ${c_n}^j$ in terms of $M_i$.
We thus obtain the $M_i$ dependent truncated wavefunction, say, 
$\Phi(x;M_i)$. We can calculate a new mass eigenvalue $M_{i+1}$ using 
this wavefunction as 
\beq
M_{i+1}^2={<\Phi(M_i)|H|\Phi(M_i)>\over <\Phi(M_i)|\Phi(M_i)>}.
\endeq
We can use Eq.(\ref{aprox}) as $M_0^2$. For $N\le 15$ and $m=0.01$, 
mass $M^2$ converges in 5 iterations. For $0<m<0.5$, we obtain 
$M^2$'s which are summarized in Table \ref{table2}. We can fit them 
by polynomials, as follows:
\beq
M^2(\alpha=0,m)&=&3.62763m + 3.58027{m^2} + 0.0683573{m^3} +O(m^4)\nn
M^2(\alpha=1,m)&=&1+ 3.62421m + 3.34492{m^2} + 0.213839{m^3}+O(m^4).
\endeq
It should be noted here that the coefficients of $m$ are consistent 
with Eq.(\ref{aprox}) and Bergknoff's result.  In order to see the 
efficacy of this new basis function expansion, we show the 
wavefunctions in Fig.~3.

\fbox{Table 2}

\fbox{Fig.~3}

\section{Summary and Discussion}
In the preceding sections we have introduced the new basis function
and calculated the mass eigenvalue of the bound state using the new 
basis function. We have found that 
 (1) the new basis function gives an effective 
approximation of the wavefunction, and (2) the mass eigenvalues 
are consistent with the results of the precursors. 
In the remainder of this section, we will discuss 
the 2\% discrepancy problem.

Let us consider the wavefunction given by
\beq
\Phi(x)=(1-4x^2)^{\gamma_0} +a_1 (1-4x^2)^{\gamma_1},
\endeq
where we assume only $\gamma_0 = {\gamma_0}^1 m+ {\gamma_1}^3 m^3 
+O(m^5)$, $\gamma_1 = {\gamma_1}^0 + {\gamma_1}^1 m +O(m^2)$, and 
$a_1={a_1}^1 m^{1/2+\delta} +O(m)$. We are then led to
\beq
M^2\equiv{<\Phi|H|\Phi>/<\Phi|\Phi>}=
\alpha +\left(
{\frac{1}{{\gamma_0}^1}} + {\frac{{\gamma_0}^1 {{\pi }^2}}{3}}\right)m
+O(m^{1+2\delta}).
\label{universal}
\endeq
This relation holds independently of the details of $\gamma_0$, 
$\gamma_1$, and $a_1$, except for certain assumptions which were made 
before Eq.(\ref{universal}). The coefficient of $m$ in 
Eq.(\ref{universal}) has the minimum value ${2\pi\over \sqrt{3}}$ 
when ${\gamma_0}^1={\beta_0}^1\equiv {\sqrt{3}\over \pi}$. 
We may therefore conclude that Eq.(\ref{universal}) holds universally, 
provided that:\\
 (1) The wavefunction can be expanded as a power series of 
$(1-4x^2)$, like \\
\indent $\Phi(x)=(1-4x^2)^{\gamma_0}+\sum_{j=1}^{\infty}a_j 
(1-4x^2)^{\gamma_j}$, $\gamma_0<\gamma_1<\cdots$,\\
and\\
(2)The coefficients of the series, $a_j$'s, are of order $m^{1/2+\delta}$.

An almost identical result has been obtained by Harada {\it et al} 
\cite{harada98}, in which they have restricted themselves to a case 
where $\gamma_0=\beta_0$, and $\gamma_j=\beta_0+j$. 
Our conclusion is a generalization of Harada {\it et al}'s result. 
We, therefore, cannot solve the ``2\% discrepancy'' problem in the 
context of the 't~Hooft-Bergknoff-Eller equations. 

\section*{Acknowledgements}
The author would like to thank Professor K. Tanaka and Professor 
G.J. Aubrecht for comments and discussions during the early stage of
this work. He is also grateful to Dr. Harada for useful discussions of
the 2\% discrepancy problem and related topics.
This work was supported by the Grants-in-Aid for Scientific Research of 
Ministry of Education, Science and Culture of Japan (No. 10640198).
\appendix
\section{Proof of Eq.(3)}
\label{ap1}
We will prove Eq.(\ref{eq1}). For monomial $x^n$, 
by the use of the definition of the finite part integral, we obtain
\beq
\L :x^n\mapsto -{4x^n\over 1-4x^2}+ n x^{n-1}\log{1-2x\over 1+2x}
+f_n(x).
\endeq
Here,
\beq
f_m(x)&=&\sum_{k=2}^m {_mC_k\over k-1}\left\{ 
\left(\half -x\right)^{k-1}-\left(-\half -x\right)^{k-1}
\right\}x^{m-k}\nn
&=&\cases{
  {\displaystyle \sum_{k=0}^{n-1}{(2k+1)x^{2k}\over 
      \{2(n-k)-1\}2^{2n-2k-2}  }},
  \;{\rm for}\; m=2n,
  \cr
  {\displaystyle \sum_{k=0}^{n-1}{(2k+2)x^{2k+1}\over 
      \{2(n-k)-1\}2^{2n-2k-2}}},
  \;{\rm for}\; m=2n+1.
  \cr}
\endeq
\par
Using identities
\beq
\suma{n}c_{2n}\left(-\frac{4x^{2n}}{1-4x^2}\right)
  =-\suma{n}\sum_{k=0}^nc_{2k}4^{n-k+1}x^{2n},
\endeq
\beq
\suma{n} 2n\, c_{2n}x^{2n-1}\log{1-2x\over 1+2x}
=-\suma{n}\left[\sum_{k=0}^{n}c_{2k}{2^{2n-2k+3}k\over 
   2(n-k)+1}\right]x^{2n},
\endeq
and
\beq
\sum_{n=1}^{\infty}c_{2n}f_{2n}(x)=-\suma{n}\left[
   \sum_{k=0}^{n}c_{2k}{(2n+1)2^{2n-2k+1}\over 2(n-k)+1}\right]x^{2n},
\endeq
we see that, for a given even function $\suma{n}c_{2n}x^{2n}$,
\beq
\L &:&\suma{n} c_{2n} x^{2n}\mapsto
-\suma{n}(2n+1)\suma{k}{4^{n-k+1}c_{2k}\over 2(n-k)+1}x^{2n}.
\label{expandorg}
\endeq
\par
Power series expansion
\beq
(1-4x^2)^\beta =\suma{n} {(-\beta )_n(4x^2)^n\over n!},
\endeq
holds where $\displaystyle (a)_n\equiv {\Gamma (a+n)\over \Gamma (a)}$ 
is the Pochhammer symbol. Thus, for $\displaystyle\, c_{2n}=
{(-\beta )_n4^n\over n!}$ and $c_{2n+1}=0$, we see that
\beq
\L : (1-4x^2)^\beta \mapsto
-{4\pi^{1/2}\Gamma(1+\beta)\over \Gamma(1/2+\beta)}
  F(1,1/2-\beta;1/2;4x^2),
\endeq
where $F(a,b;c;z)$ is the Gauss' hypergeometric function.
Analogously, we have 
\beq
\L : x(1-4x^2)^\beta \mapsto
-{8\pi^{1/2}\Gamma(1+\beta)\over \Gamma(1/2+\beta)}x\,
 F(2,1/2-\beta;3/2;4x^2).
\endeq

\begin{table}
\caption{Number of nontrivial equations}
\label{table1}
\begin{center}
\begin{tabular}{cccc}
   terms                    & range of n & range of $j$   & 
                 number of nontrivial equations \\
\hline
$(1-4x^2)^{\beta_n-1}$    & $0\sim N$    &  ---         & 
                $0$ (automatically satisfied) \\
\hline
$(1-4x^2)^{\beta_n+j}$  & $0\sim N$    &  $0\sim N-n-1$ & 
                $\frac{N(N+1)}{2}$           \\
\hline
$(1-4x^2)^{j}$              &    ---       &  $0\sim N-1$   &  
                $N$                          \\
\end{tabular}
\end{center}
\end{table}
\begin{table}
\caption{Numerical results for bound state mass $M^2$ in $\rm SU(N_C)$ 
and $\rm U(N_C)$ models as a function of quark mass $m$}
\label{table2}
\begin{center}
\begin{tabular}{ccccccc}
    m              & 0.01     & 0.10     & 0.20     & 0.30     
        & 0.40     & 0.50  \\
\hline
$M^2$ in $\rm SU(N_C)$ & 0.036634 & 0.398634 & 0.869282 & 1.412358 
        & 2.028271 & 2.717415\\
\hline
$M^2$ in $\rm U(N_C)$  & 1.036607 & 1.396177 & 1.860377 & 2.394130 
        & 2.998685 & 3.675073\\
\end{tabular}
\end{center}
\end{table}
\begin{figure}
\vspace*{-4cm}
\begin{center}
\includegraphics[width=\graphicwidth]{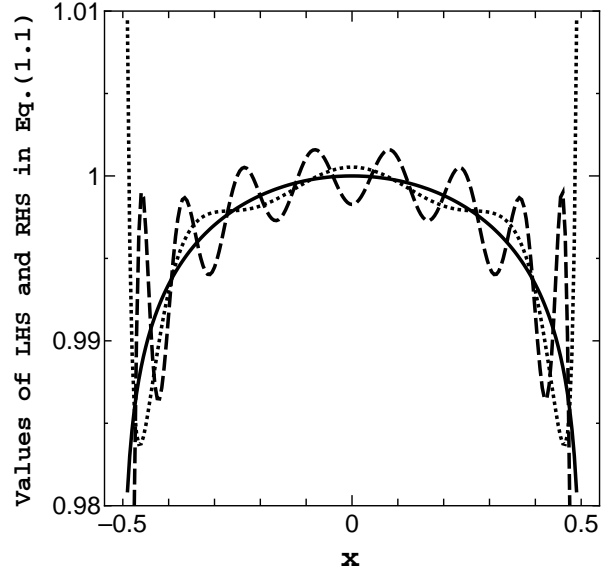}
\vspace*{-1cm}
\caption{The comparison of the relative values of both sides of 
Eq.(\protect{\ref{eller}}) for $\alpha=0$. The wavefunction in 
Eq.(\protect{\ref{eller}}) was approximated by 
Eq.(\protect{\ref{expansion}}) with $N=3$ and 
Eq.(\protect{\ref{numeric}}).
The solid line represents the LHS and the dotted line stands for the 
RHS. The RHS with $N=9$, which is indicated by the dashed line, is 
exhibited for comparison.}
\end{center}
\end{figure}
\vspace*{-2.cm}
\begin{figure}
\begin{center}
\includegraphics[width=\graphicwidth]{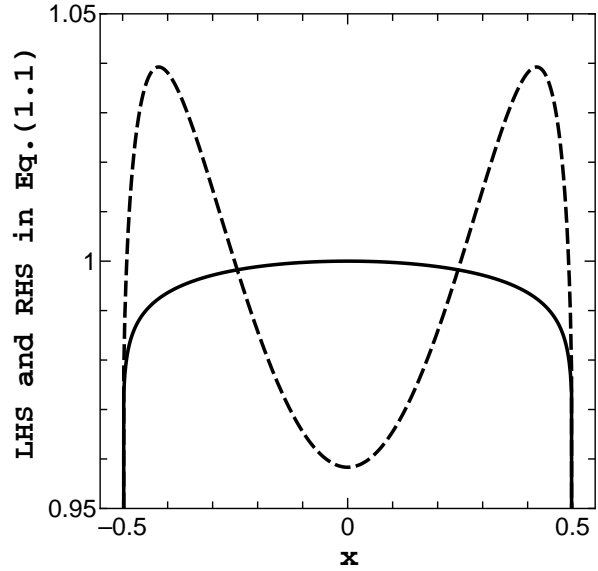}
\vspace*{-1cm}
\caption{The comparison of the relative values of both sides of 
Eq.(\protect{\ref{eller}}) for $\alpha=0$. The wavefunction in
Eq.(\protect{\ref{eller}}) was approximated by 
Eq.(\protect{\ref{newwave}}) with $N=1$ and 
Eq.(\protect{\ref{aprox}}) with $m=0.01$.
The solid line represents the LHS and the dashed line stands for the 
RHS.}
\end{center}
\end{figure}
\begin{figure}
\begin{center}
\includegraphics[width=\graphicwidth]{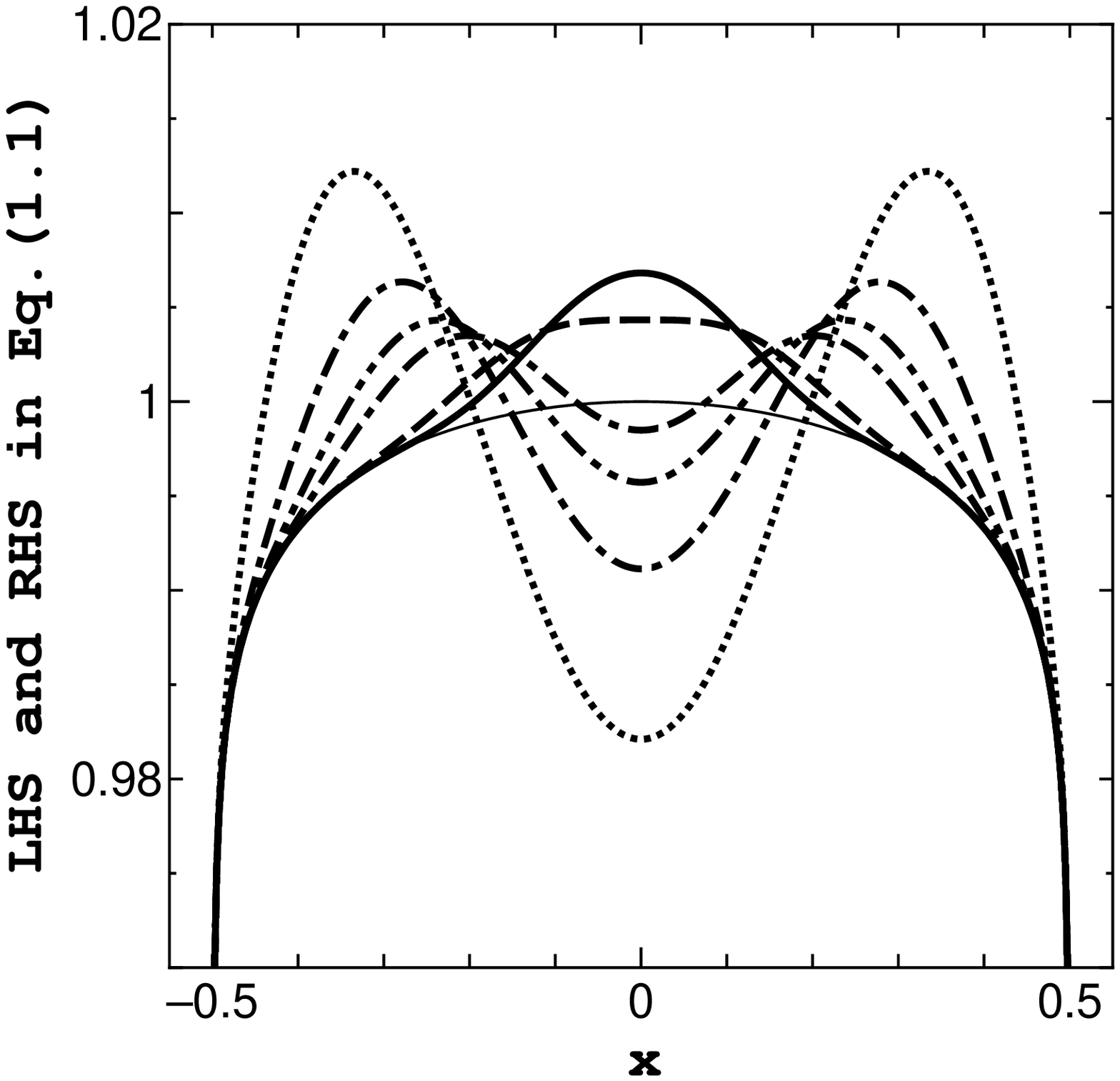}
\vspace*{-1cm}
\caption{The convergence of the new basis function expansion 
for $m=0.01$ fixed. The thin solid line represents the LHS in 
Eq.(\protect{\ref{eller}}), provided that the wavefunction was 
approximated by
Eq.(\protect{\ref{newwave}}) with $N=15$. The dotted line denotes
the RHS with $N=2$, the dot-dashed line exhibits the RHS with $N=3$, 
the dot-dot-dashed line represents the RHS with $N=4$, 
the dot-dash-dashed line stands for the RHS with $N=5$, and the dashed 
line exhibits the RHS with $N=10$. The thick solid 
line indicates the RHS in Eq.(\protect{\ref{eller}}) with wavefunction 
given in Eq.(\protect{\ref{newwave}}) with $N=15$.}
\end{center}
\end{figure}

\begin{references}
\bibitem{brodsky98}
S. J. Brodsky, H-C. Pauli, and S. Pinsky,
Phys. Rep. {\bf 301}, 299 (1998).
\bibitem{burkardt}
M. Burkardt, 
Adv. Nucl. Phys. {\bf 23} 1 (1996).
\bibitem{tammdancoff}
I. Tamm, J. Phys. (Moskow) {\bf 9}, 449 (1945);
S. M. Dancoff, Phys. Rev. {\bf 78}, 382 (1950).
\bibitem{lftd}
R. J. Perry, A. Harindranath, and K. G. Wilson, 
Phys. Rev. Lett. {\bf 65}, 2959 (1990).
\bibitem{ma}
Y. Ma and J. R. Hiller,
J. Comp. Phys. {\bf 82}, 229 (1989).
\bibitem{mo}
Y. Mo and R. J. Perry,
J. Comp. Phys. {\bf 108}, 159 (1993).
\bibitem{harada94}
K. Harada, T. Sugihara M. Taniguchi and M. Yahiro
Phys. Rev. {\bf D49}, 4226 (1994).
\bibitem{coleman}
S. Coleman, R. Jackiw, and L. Susskind, 
Ann. Phys. (N.Y.) {\bf 93}, 267 (1975);
S. Coleman, 
Ann. Phys. (N.Y.), {\bf 101}, 239 (1976).
\bibitem{schwinger}
J. Schwinger, 
Phys. Rev. {\bf 128}, 2425 (1962).
\bibitem{bergknoff}
H. Bergknoff, Nucl. Phys. {\bf B122}, 215 (1977). 
\bibitem{sugihara}
T. Sugihara, M. Matsuzaki and M. Yahiro,
Phys. Rev. {\bf D50}, 5274 (1994).
\bibitem{thooft}
G. 't~Hooft, Nucl. Phys. {\bf B75}, 461 (1974).
\bibitem{hornbostel}
K. Hornbostel, S. J. Brodsky and H.-C. Pauli
Phys. Rev. {\bf D41}, 3814 (1990).
\bibitem{banks}
T. Banks, J. Kogut, and L. Susskind,
Phys. Rev. {\bf D13}, 1043 (1976).
\bibitem{harada98}
K. Harada, T. Heinzl, and C. Stern
Phys. Rev. {\bf D57}, 2460 (1998).
\bibitem{huang}
S. Huang, J.W. Negele and J. Polonyi,
Nucl. Phys. {\bf B307}, 669 (1988).
\bibitem{sande}
B. van de Sande, 
Phys. Rev. {\bf D54}, 6437 (1996).
\bibitem{duncan}
A. Duncan, S.Pernice, and E. Schnapka, 
Phys. Rev. {\bf D55}, 2422 (1997).
\bibitem{abe}
O. Abe, G. J. Aubrecht, and K. Tanaka,
Phys. Rev. {\bf D56}, 2242 (1997).
\bibitem{abramowitz}
M. Abramowitz and I. A. Stegun,
{\it Handbook of Mathematical Functions with Formulas, Graphs, and 
Mathematical Tables},
(Dover Publ., Inc., New York, 1972).
\end{references}
\end{document}